\PassOptionsToPackage{table}{xcolor}
\documentclass[letterpaper,twocolumn,10pt]{article}
\usepackage{group}

\usepackage{times}
\usepackage{latexsym}

\usepackage[T1]{fontenc}
\usepackage{microtype}
\usepackage{inconsolata}
\usepackage{graphicx}
\usepackage{pifont}
\usepackage{booktabs}
\usepackage{xspace}
\usepackage{enumitem}
\usepackage{tabularx}
\usepackage{xcolor}
\usepackage[most]{tcolorbox}
\usepackage{hyperref}
\usepackage{cleveref}
\usepackage{algorithm}
\usepackage{algpseudocode}
\usepackage{booktabs}
\usepackage{multirow}
\usepackage{algorithm}
\usepackage{algpseudocode}
\usepackage{booktabs}

\definecolor{privacyboxbg}{HTML}{F3F7F6}
\definecolor{privacyboxframe}{HTML}{9DB7AE}
\definecolor{privacyboxtitle}{HTML}{3F6258}

\newtcolorbox{privacytagbox}[1]{
    enhanced,
    colback=privacyboxbg,
    colframe=privacyboxframe,
    coltitle=privacyboxtitle,
    fonttitle=\bfseries,
    title=#1,
    boxrule=0.6pt,
    arc=2mm,
    left=.5mm,
    right=.5mm,
    top=.2mm,
    bottom=.2mm
}
\definecolor{takeawayboxbg}{HTML}{EDF6F2}
\definecolor{takeawayboxframe}{HTML}{7FAF9A}
\definecolor{takeawayboxtitle}{HTML}{365F50}

\newtcolorbox{takeawaybox}{
    enhanced,
    colback=takeawayboxbg,
    frame hidden,
    borderline west={1.9pt}{0pt}{takeawayboxframe},
    boxrule=0pt,
    arc=1mm,
    left=1mm,
    right=0.6mm,
    top=.6mm,
    bottom=.6mm,
    boxsep=0.6mm,
    before skip=4pt,
    after skip=8pt
}

\definecolor{promptboxbg}{HTML}{F4F7FB}
\definecolor{promptboxframe}{HTML}{8EA9C7}
\definecolor{promptboxtitle}{HTML}{2F4F6F}

\newtcolorbox{promptbox}[1]{
    enhanced,
    breakable,
    colback=promptboxbg,
    colframe=promptboxframe,
    coltitle=promptboxtitle,
    fonttitle=\bfseries,
    title=#1,
    boxrule=0.6pt,
    arc=1mm,
    left=0.5mm,
    right=0.5mm,
    top=.5mm,
    bottom=.5mm,
    boxsep=0.5mm,
    before skip=2pt,
    after skip=2pt
}

\newcommand{\mypara}[1]{\noindent{\bf {#1}.}\xspace}

\definecolor{algblue}{HTML}{1F6FAE}
\DeclareRobustCommand{\algjump}[2]{
  \kern0.2em
  \hyperref[#1]{\textcolor{algblue}{\scriptsize$\triangleright$~#2 \S\ref*{#1}}}
}

\begin{document}

\title{UniCA: Bi-directional Cross-Attention with Positive Similarity Loss for Robust Multi-Modal Retrieval}
\date{}

\author{
Yini Huang\textsuperscript{1} \quad
Wenlong Zhang\textsuperscript{2} \\[0.4cm] 
\textsuperscript{1}\textit{The Hong Kong University of Science and Technology (Guangzhou), Guangdong, China} \\
\textsuperscript{2}\textit{Southern Medical University, Guangdong, China}
}

\maketitle

\begin{abstract}
\hspace{\parindent}Multi-modal retrieval has become increasingly critical for handling the growing volume of integrated visual-textual data in real-world applications, but existing frameworks rely on implicit fusion via text encoder self-attention, limiting explicit cross-modal semantic alignment.
To address this gap, this paper proposes UniCA (Unified Cross-Attention Encoder), a multi-modal retrieval model with four key innovations:
1) a bi-directional cross-attention (Bi-CA) block that enables active semantic exchange between visual and textual tokens prior to concatenation, capturing inter-modal correlations more efficiently.
2) a Positive Similarity Loss that optimizes absolute semantic proximity between query and positive candidate embeddings.
3) a streamlined dataset UMR-S10 (Universal Multimodal Retrieval Sample 10\%) to reduce computational costs while retaining semantic diversity and task representativeness.
4) an experimental validation on the WebQA benchmark demonstrates that UniCA outperforms the baseline model across Hybrid and Image-Text tasks, achieving improvements of up to 4.09\% in Recall@5, 3.28\% in Recall@10, and 3.96\% in MRR@1 for the hybrid task.
UniCA provides an efficient and robust solution for multi-modal retrieval, lowering deployment barriers through its lightweight dataset and enhanced fusion mechanism.
\end{abstract}

\section{Introduction}

In the era of information explosion, Information retrieval (IR) serves as a foundational component in a wide range of real-world applications.
Traditional dense retrieval models, such as DPR \cite{karpukhin2020dense} , Contriever \cite{izacard2021unsupervised} , and BGE \cite{xiao2024c}, have demonstrated remarkable performance in text-based retrieval tasks.
However, these models are inherently limited to processing textual data, failing to accommodate the increasingly prevalent multi-modal information that integrates both visual and textual content.

Multi-modal retrieval, a critical extension of traditional IR, aims to retrieve semantically relevant results across different data modalities.
It has attracted extensive attention from both academic and industrial communities due to its broad application prospects in scenarios.
The core challenge of multi-modal retrieval lies in learning effective unified representations that align semantic information across diverse modalities, thereby enabling accurate matching between query and candidate results of different modal types.
The emergence of vision-language models (VLMs) has significantly advanced cross-modal understanding, laying a solid foundation for multi-modal retrieval.
Early works like LXMERT \cite{tan2019lxmert} use cross-modality encoder to achieve in-depth interaction of visual and language information.
Subsequently, CLIP \cite{radford2021learning} demonstrated that contrastive pre-training on massive internet-scale image-text pairs could learn powerful and semantically aligned representations, which have been widely adapted for multi-modal retrieval due to their strong transferability. 

To advance the performance of multi-modal IR towards universal applicability, the research community has shifted its focus to developing large-scale universal multi-modal models.
Among these efforts, the VISTA \cite{zhou2024vista} framework stands out as a promising approach tailored for universal multi-modal retrieval.
It mitigates modality imbalance by projecting one modality directly into the representation space of another.
However, its architecture relies solely on the subsequent Text Encoder's self-attention to implicitly handle the deep fusion of concatenated visual and textual tokens.
This reliance can limit the model's ability to swiftly and explicitly establish cross-modal semantic connections.

In this work, we propose UniCA, an Unified Cross-Attention Encoder model, to address the above issues.
The four key contributions are:

Firstly, this paper introduces bi-directional Cross-Attention layers between the visual and text tokens prior to concatenation.
This forces an active, semantic exchange at the input level, enhancing the efficiency and quality of the final multi-modal representation.

Secondly, this paper proposes a contrastive loss function not only to maintain the category discrimination ability of the model but also enhance the semantic alignment between query and positive candidate embeddings.

Thirdly, this paper constructs a streamlined dataset UMR-S10 (Universal Multimodal Retrieval Sample 10\%) by sampling 10\% of the core samples from original VISTA \cite{zhou2024vista} dataset.
This lightweight dataset significantly reduces training time and computational resources, effectively lowering deployment barrier while preserving model performance.

Fourthly, extensive experiments on the WebQA validate UniCA’s superiority over the baseline.
UniCA achieves 4.09\% improvement in Recall@5, 3.28\% improvement in Recall@10, and 3.96\% improvement in MRR@1 on the Hybrid task, demonstrating the effectiveness of our proposed modules in advancing multi-modal retrieval performance.

\begin{figure}[t]
    \centering
    \includegraphics[width=\linewidth]{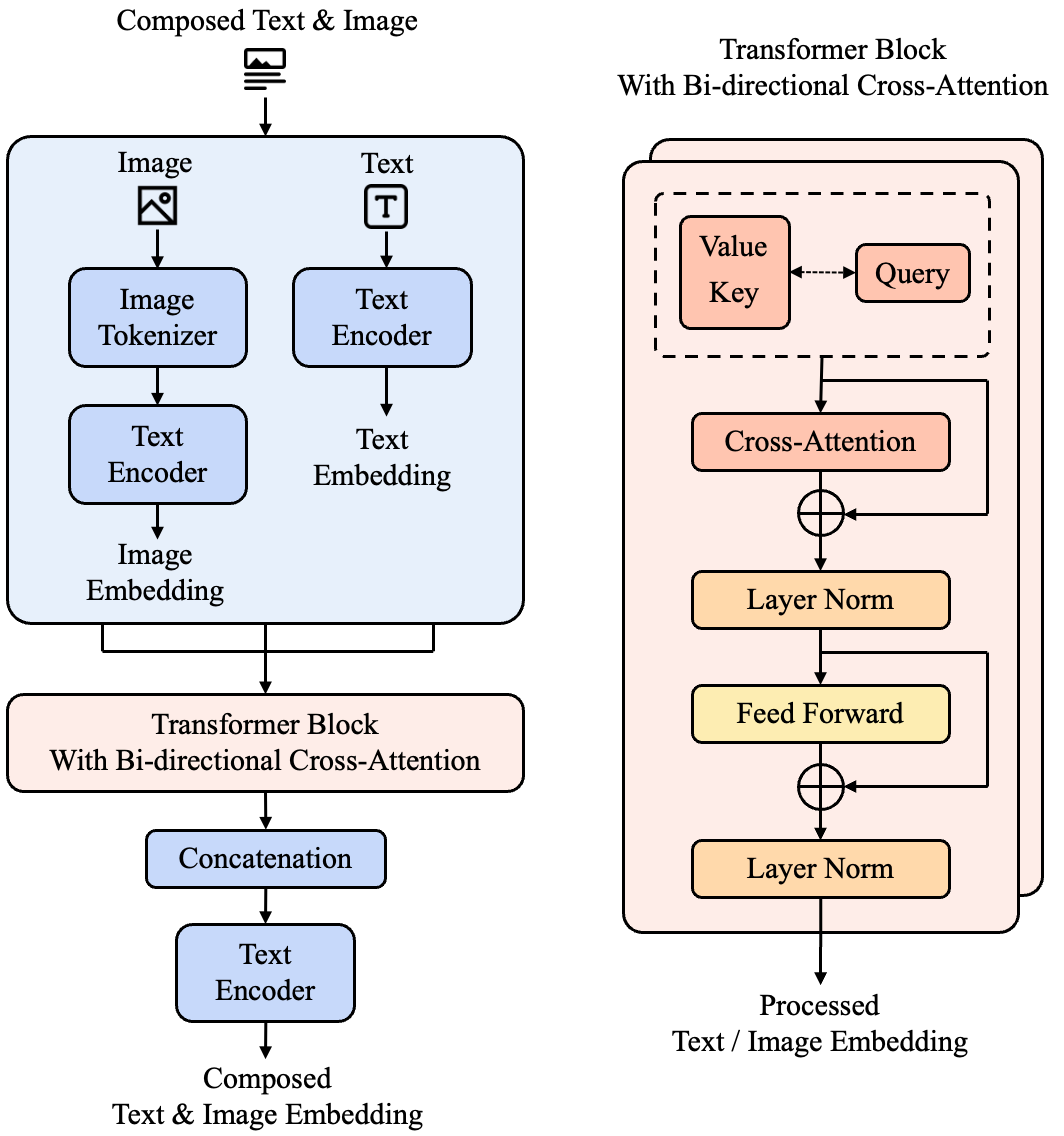}
    \caption{The Architecture of UniCA.}
    \label{fig:overview_UniCA}
\end{figure}

\section{Related Work}

Dense Retrieval (DR) has become the dominant paradigm in modern Information Retrieval (IR).
The foundation of this success lies in the development of highly effective general text embedding models, which map queries and documents into a low-dimensional vector space where semantic relevance is captured by vector proximity.
Early models utilized BERT \cite{devlin2019bert} fine-tuned via contrastive learning, leading to models like DPR \cite{karpukhin2020dense} and later iterative advancements such as Contriever \cite{izacard2021unsupervised}, GTR \cite{ni2022large}, and E5 \cite{wang2022text}.
Recent state-of-the-art text-only models, like BGE \cite{xiao2024c}, continue to push the boundaries of text representation quality and generalization.
While powerful, these models are strictly text-oriented, incapable of handling the growing volume of visual and composite image-text data found in real-world scenarios, which necessitates the expansion to multi-modal capabilities.

The field of Vision-Language Pretraining (VLP) aims to learn robust joint representations between visual and textual data, which is critical for extending retrieval capabilities to multi-modal domains.
Early research, such as LXMERT \cite{tan2019lxmert} introduced cross-encoder architectures to enable fine-grained interactive fusion of visual and linguistic features, supporting complex cross-modal reasoning.
ViT \cite{dosovitskiy2020image} revolutionized visual modeling by splitting images into fixed-size patches, embedding these patches into sequences, and processing them through a Transformer encoder.
This design not only enabled direct compatibility with text Transformer architectures but also provided a flexible framework for integrating visual and linguistic information.
Building on ViT \cite{dosovitskiy2020image}, MAE \cite{he2022masked} further enhanced visual representation learning through self-supervised masked patch reconstruction, allowing models to capture more robust and generalizable visual features with fewer labeled data, which in turn boosted the performance of downstream VLP tasks.
In parallel, CLIP \cite{radford2021learning} leveraged contrastive pre-training on massive image-text pairs to learn semantically aligned representations, enabling powerful zero-shot cross-modal retrieval and setting a benchmark for multi-modal IR.
To address the heavy computational burden of traditional models, ViLT \cite{kim2021vilt} introduced a convolution-free minimalist design that simplifies the visual embedding pipeline, significantly reducing inference costs while maintaining competitive performance. Recent efforts focus on balancing the accuracy of cross-encoders with the efficiency of dual-encoders.
The BLIP \cite{li2022blip} series proposed unified frameworks that efficiently leverage frozen, powerful pre-trained components.
Notably, BLIP-2 \cite{li2023blip} introduced the Q-Former—a lightweight querying Transformer—to bridge the modality gap between frozen visual encoders and Large Language Models (LLMs), achieving a synergistic improvement in both performance and computational efficiency.

Similarly, the VISTA \cite{zhou2024vista} framework operates as a visual plugin for a robust text encoder , mitigating modality imbalance by mapping visual tokens into the Text Encoder's input space.
While VISTA \cite{zhou2024vista} is engineered to generate general multi-modal embeddings , its fusion mechanism relies on the implicit capability of the subsequent Text Encoder's self-attention to interpret the concatenated visual and textual tokens.
This reliance on passive fusion can limit the model's ability to swiftly and explicitly establish fine-grained semantic connections.
UniCA model is specifically designed to overcome this passive fusion bottleneck in the VISTA architecture.

\section{Methodology}

This paper proposes UniCA, an Unified Cross-Attention Encoder model, maintaining the core philosophy of a dual-encoder architecture designed for universal multi-modal retrieval.
UniCA retains the structural foundation in VISTA \cite{zhou2024vista} but introduces critical enhancements in the fusion mechanism, the training objective, and the resource strategy.

\subsection{Model Architecture }

The architecture of UniCA is built upon the principle of preserving text retrieval capability while enhancing cross-modal fusion.
This method ensures the model can not only maintain the compatibility with existing text-intensive retrieval tasks, but also adapt to multi-modal scenarios.

As shown in \cref{fig:overview_UniCA}, this diagram shows the model pipeline: image/text inputs are encoded into aligned embeddings, processed via a bi-directional cross-attention block, then concatenated into a unified multi-modal representation for retrieval.

UniCA facilitates a unified embedding space by processing different data modalities through tailored pipelines.

For text-only data, the pipeline maintains a standard dense retrieval flow. Raw text is directly fed into the frozen Text Encoder, which outputs a normalized [CLS] token embedding as the final text representation.
This retains the Text Encoder’s strong textual semantic capabilities, ensuring compatibility with traditional text retrieval systems.

For image-only data, the Image Tokenizer first splits the image into patches and converts them into visual tokens.
These tokens are then projected into the Text Encoder’s semantic space and fed into the Text Encoder with a prepended [CLS] token to generate a normalized image embedding—aligning visual features with the text space for cross-modal matching.

For composed text \& image data, the two modalities are processed separately first.
The image is encoded into an image embedding via Image Tokenizer and Text Encoder as mentioned above, and the text is encoded into a text embedding by a Text Encoder.
These two embeddings are then passed through the Bi-directional Cross-Attention Block: text acts as Query to attend to image features, and image acts as Query to attend to text features, enabling explicit cross-modal interaction.
Finally, the enhanced embeddings are concatenated and encoded by the Text Encoder to produce a unified multi-modal embedding, capturing the combined semantics of both modalities.

UniCA consists of three core components: a frozen pre-trained Text Encoder, a trainable ViT-based Image Tokenizer, and a Bi-directional Cross-Attention Block.

\mypara{Text Encoder}
UniCA adopts a pre-trained general text embedding model as the Text Encoder, which is fully frozen during all training stages.
This design is motivated by two key considerations.
First, state-of-the-art text encoders have demonstrated superior performance in text-only retrieval tasks, and freezing them ensures UniCA retains this strong text representation capability—critical for scenarios where multi-modal data coexists with text-only data.
Second, freezing the Text Encoder reduces the number of trainable parameters, lowering computational costs.
The Text Encoder’s architecture follows the standard BERT \cite{devlin2019bert} design: 12 transformer layers, 12 attention heads, and a hidden dimension of 768.
For embedding output, this paper uses the normalized hidden state of the [CLS] token—consistent with mainstream text embedding practices—to ensure compatibility with dense retrieval pipelines.

\mypara{Image Tokenizer}
The Image Tokenizer is the critical bridge that translates raw visual signals into the linguistic format required by the Text Encoder.
In UniCA, this module is based on a Vision Transformer (ViT) \cite{dosovitskiy2020image} and is initialized from the baseline model.
The tokenization process involves several sophisticated stages.
The input image is first divided into a sequence of non-overlapping flattened patches, which are mapped through a trainable linear projection to create initial “visual words”.
To preserve structural layout, learnable spatial positional embeddings are added to the patch tokens, enabling the model to understand the relative positions of objects.
These tokens then pass through a series of Transformer blocks to capture long-range dependencies across the image.
Finally, a linear projection layer is applied to the hidden states to act as a modality translator, mapping visual features directly into the embedding space of the Text Encoder.
By training this module, UniCA effectively transforms visual data into tokens that the frozen Text Encoder can interpret as high-level semantic concepts.

\subsection{Bi-directional Cross-Attention}
The most critical innovation of UniCA is the bi-directional cross-attention (Bi-CA) transformer block, introduced to enable explicit cross-modal fusion before the concatenation of visual and textual tokens.
The Bi-CA Block consists of two cross-attention sub-blocks (Text-to-Image Cross-Attention and Image-to-Text Cross-Attention) and a shared feed-forward network with layer normalization.
The Bi-CA Block’s two sub-blocks operate in parallel, which ensure the bidirectional information flow between modalities. The detailed structure of a single cross-attention sub-block is as follows.

\mypara{Cross Attention Block}
Each cross-attention sub-block follows the standard transformer cross-attention design \cite{vaswani2017attention}, but enforces strict separation of Query (Q) and Key/Value (K/V) sources.
This separation is what distinguishes cross-attention from self-attention (where Q, K, and V derive from the same sequence) and enables explicit cross-modal interaction.

When processing a text sequence as Q and a visual sequence as K and V, the sub-block first projects each sequence into three distinct subspaces by learnable linear layers.
This projection step transforms the original 768-dimensional features into 64-dimensional sub-features (matching the dimension of each attention head), which are then used to compute attention scores.
These scores quantify the semantic relevance between every token in the Query sequence and every token in the Key and Value sequence.

The same mechanism applies symmetrically when an image sequence serves as Q, with a text sequence acting as K and V.
The image tokens are projected into the same set of subspaces, and attention scores are calculated to capture semantic associations between visual tokens and textual tokens.
Formally, the attention output is calculated as:

\begin{equation}
\text{Attention}(Q, K, V) = \text{softmax}\left(\frac{QK^T}{\sqrt{d_k}}\right)V
\tag{1}
\end{equation}

\mypara{Layer Normalization and Residual Connection}
After the cross-attention computation, the output is combined with the original Query sequence by a residual connection, a design choice that mitigates gradient vanishing during training by preserving the original Query’s semantic information while adding the cross-modal enhancement.
This combined output is then normalized using LayerNorm, which standardizes the feature distribution across batches and layers.
Together, these steps ensure the cross-modal features remain stable and expressive, even as the model trains on multi-modal data with varying semantic distributions.

\mypara{Feed-Forward Network (FFN)}
The FFN introduces non-linearity to the cross-modal features, enabling the model to capture complex semantic relationships.
Its computation unfolds in three sequential steps.
First, the 768-dimensional normalized cross-attention output is passed through a linear layer to expand its dimension to 3072 (four times the original hidden size).
This expansion increases the model’s capacity to capture fine-grained semantic patterns.

\begin{equation}
x_1 = \text{Linear}_{768 \to 3072}(x)
\tag{2}
\end{equation}

The expanded feature ($x_1$) is then processed with a GELU activation function, which introduces non-linearity to the feature space---critical for modeling non-trivial relationships between text and image signals.

\begin{equation}
x_2 = \text{GELU}(x_1)
\tag{3}
\end{equation}

Finally, a second linear layer compresses the 3072-dimensional activated feature back to the original 768-dimensional space (matching the Text Encoder’s hidden dimension).

\begin{equation}
x_{out} = \text{Linear}_{768 \to 3072}(x_2)
\tag{4}
\end{equation}

\subsection{The Training Objective}
To develop robust multi-modal representation capabilities, UniCA is optimized through a specialized training objective designed to bridge the semantic gap between composite queries and their target candidates.
While traditional retrieval models, primarily rely on Cross-Entropy loss to distinguish positive pairs from in-batch negatives, this paper observes that Cross-Entropy loss focuses on relative ranking rather than absolute semantic proximity.
For multi-modal tasks, where visual and textual signals must be tightly aligned at the semantic level, this relative ranking focus can lead to suboptimal embedding geometry, especially when hard negatives are present.

To address this, UniCA introduces a paradigm shift in the training objective by directly supervising the cosine similarity between paired embeddings.

Our training objective is defined as a Positive Similarity Loss, which explicitly minimizes the semantic distance between a query representation and its corresponding positive candidate.
Formally, given a query embedding $q$ and a positive candidate embedding $c^+$, both normalized to the unit hypersphere, their semantic alignment can be calculated as the inner product:

\begin{equation}
\text{sim}(q, c^+) = q^T c^+
\tag{5}
\end{equation}

The training loss $\mathcal{L}$ is then formulated to maximize this similarity across the entire batch $\mathcal{B}$:

\begin{equation}
\mathcal{L} = 1 - \frac{1}{|\mathcal{B}|} \sum_{i \in \mathcal{B}} \text{sim}(q_i, c_i^+)
\tag{6}
\end{equation}

This objective ensures that the model not only discriminates positives from negatives but also aggressively clusters the multi-modal representations of positive pairs into the same local region of the embedding space.
By focusing on absolute semantic proximity rather than batch-wise ranking, UniCA learns a unified embedding space where visual-textual correlations are encoded as geometric closeness---critical for capturing the fine-grained semantic alignment required by complex multi-modal retrieval tasks.

\subsection{Dataset Construction}
This paper adopted the self-constructed lightweight dataset UMR-S10 (Universal Multimodal Retrieval Sample 10\%), derived from 10\% core samples of the original VISTA dataset \cite{zhou2024vista}.
The sampling strategy prioritizes data with high semantic diversity and task representativeness, ensuring the dataset retains the key characteristics of the full corpus while reducing data volume.
This streamlined dataset not only cuts down on redundant training data and reduces computational costs, but also avoids overfitting caused by excessive data.
It enables effective model training under resource-constrained conditions while ensuring that the learned multimodal alignment capability is as consistent as possible with that trained on the complete dataset.

\section{Experiments}
To verify the effectiveness of the proposed UniCA model in multimodal retrieval tasks, this section designs a series of comparative experiments and ablation studies.
The experiments are conducted on UMR-S10, focusing on evaluating the performance of UniCA in image-text and combined modality retrieval tasks.
This section details the experimental setup, results, and analysis, and quantifies the role of each innovation point through ablation studies.

\subsection{Experimental Setup}
\mypara{Settings}
This paper employed BGE-base-v1.5 \cite{xiao2024c} as the fixed Text Encoder, which consists of 12 Transformer layers with a hidden dimension of 768.
This ensures the model retains state-of-the-art text retrieval capabilities. The visual component utilizes the EVA02-CLIP-B-16 \cite{radford2021learning} as the Image Tokenizer.
To establish a rigorous and resource-efficient baseline, this paper first fine-tuned the original model by using UMR-S10. This fine-tuned version serves as our baseline model, representing the performance of the standard VISTA \cite{zhou2024vista} architecture under low-data regimes. 

For training, this paper used the AdamW optimizer with $\beta_1=0.9$, $\beta_2=0.999$, $\varepsilon=1\text{e-}8$, and a weight decay of $0.0$ (no L2 regularization). The UniCA adopts a per-device training batch size of $32$ and a learning rate of $2\text{e-}5$ (details in Sensitivity Analysis).
This paper enabled gradient checkpointing and mixed-precision training to optimize memory efficiency, with a maximum gradient norm of $1.0$ for gradient clipping.

\mypara{Benchmark and Metrics}
This paper used WebQA \cite{chang2022webqa} as the core evaluation benchmark.
This benchmark is designed to simulate the heterogeneous information reasoning process in real web search scenarios, and its core characteristics are highly compatible with the multimodal retrieval task of this experiment.
Based on the evaluation criteria for multimodal information matching and reasoning capabilities in WebQA, this experiment adopts the commonly used evaluation indicators in the field of information retrieval: Recall@5, Recall@10 and MRR@1.

\subsection{Results and Analysis}

\begin{figure*}[t]
    \centering
    \includegraphics[width=0.9\textwidth]{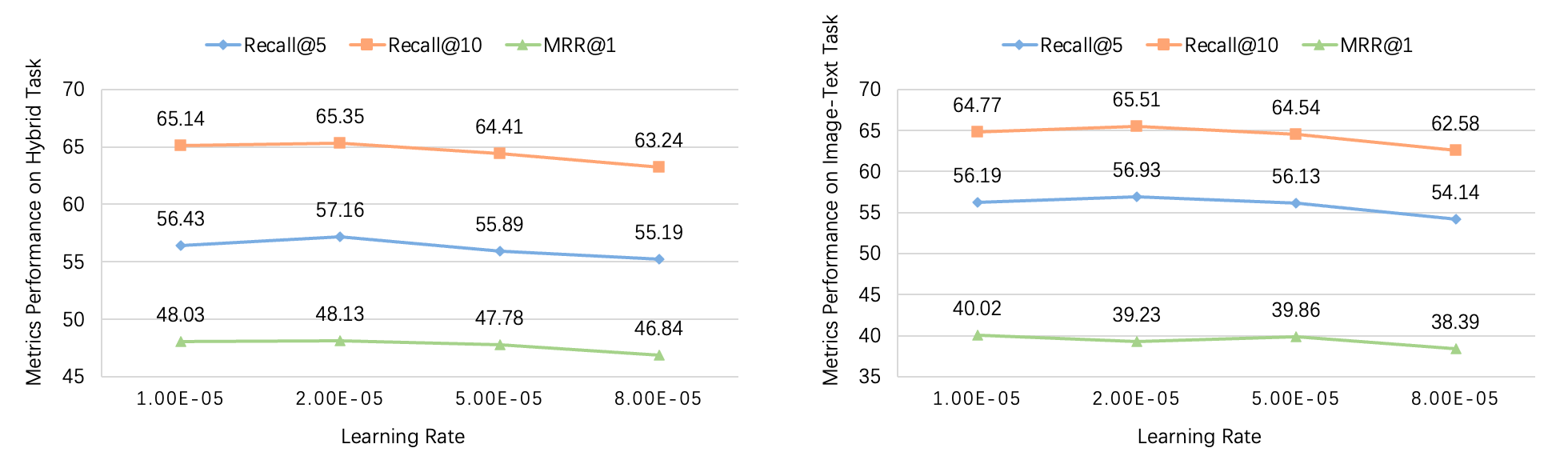}
    \caption{Sensitivity Analysis of UniCA to Learning Rate on Hybrid Tasks (left) and Image-Text Tasks (right).}
    \label{fig:sensitivity_analysis}
\end{figure*}

The main performance comparison between the proposed UniCA model and the VISTA-based baseline is summarized in \cref{tab:main_performance}.
Overall, UniCA consistently outperforms the baseline across all evaluation metrics in both the Hybrid retrieval and Image-Text retrieval tasks, demonstrating the efficacy of the bi-directional cross-attention mechanism and the similarity-based training objective.

\begin{table}[htbp]
\centering
\small

\begin{tabular}{@{}ccccc@{}}
\toprule
\textbf{Method} & \textbf{Task} & \textbf{Recall@5} & \textbf{Recall@10} & \textbf{MRR@1} \\
\midrule

\multirow{2}{*}{Baseline}
& Hybrid      & 53.07 & 62.07 & 44.16 \\
& Image-Text  & 52.27 & 61.83 & 35.92 \\
\cmidrule(r){1-5}  

\multirow{2}{*}{UniCA}
& Hybrid      & 57.16 & 65.35 & 48.12 \\
& Image-Text  & 56.83 & 65.51 & 39.23 \\

\bottomrule
\end{tabular}
\caption{Main Performance Comparison}
\label{tab:main_performance}
\end{table}

The Hybrid task, which involves complex queries composed of both image and text tokens, serves as the primary benchmark for evaluating deep cross-modal fusion.
UniCA achieves a significant performance leap, improving the Recall@5 from $53.07\%$ to $57.16\%$ and Recall@10 from $62.07\%$ to $65.35\%$. Notably, the MRR@1 sees a substantial increase of $3.96$ points from $44.16\%$ to $48.12\%$.
These improvements underscore the superiority of UniCA, allowing the model to better resolve semantic dependencies in composite queries.

In the standard Image-Text retrieval task, UniCA likewise exhibits robust gains.
Compared to the baseline, UniCA improves Recall@5 by $4.56$ points and MRR@1 by $3.31$ points. While the baseline relies on the text encoder's self-attention to implicitly align modalities, UniCA's architecture forces an active semantic exchange.
These results indicate that even when dealing with single-modality queries, the pre-fusion context provided by our cross-attention layers helps generate a more discriminative and aligned embedding space.

\subsection{Ablation Experiment}
To validate the contribution of the proposed Bi-CA module, this paper conducted an ablation experiment by removing the Bi-CA layers from UniCA (denoted as UniCA w/o Bi-CA in the table) and comparing its performance with the full UniCA model. The results are presented in \cref{tab:ablation_bi_ca}. 

\begin{table}[htbp]
\centering
\small 

\begin{tabular}{@{}ccccc@{}}
\toprule
\textbf{Method} & \textbf{Task} & \textbf{Recall@5} & \textbf{Recall@10} & \textbf{MRR@1} \\
\midrule

\multirow{2}{*}{\begin{tabular}[c]{@{}c@{}}UniCA w/o\\ Bi-CA\end{tabular}}
& Hybrid      & 48.87 & 56.68 & 41.56 \\
& Image-Text  & 38.47 & 46.95 & 26.04 \\
\cmidrule(r){1-5}

\multirow{2}{*}{UniCA}
& Hybrid      & 57.16 & 65.35 & 48.12 \\
& Image-Text  & 56.83 & 65.51 & 39.23 \\

\bottomrule
\end{tabular}
\caption{Ablation Study of Bi-CA Module}
\label{tab:ablation_bi_ca}
\end{table}

These results demonstrate that the Bi-CA module is critical for enhancing cross-modal semantic alignment: by enabling active interaction between visual and text tokens before concatenation, it effectively captures fine-grained inter-modal correlations that are missing in the UniCA w/o Bi-CA.
The larger performance gain on the Image-Text task further confirms that the Bi-CA module is particularly effective for scenarios requiring tight integration of visual and textual information.

\subsection{Sensitivity Analysis}

To explore the impact of learning rate on UniCA’s performance, this paper conducted a sensitivity analysis by evaluating four candidate learning rates on both Hybrid and Image-Text tasks, with results visualized in the figure.

As showen in \cref{fig:sensitivity_analysis}, this figure visualizes the performance of UniCA (across Recall@5, Recall@10, and MRR@1) under four candidate learning rates (1e-5, 2e-5, 5e-5, 8e-5) for the Hybrid (left) and Image-Text (right) tasks.

For the Hybrid,when the learning rate is set to 2e-5, all metrics reach their peaks.
As the learning rate increases to 5e-5 and 8e-5, they show a gradual downward trend. For the Image-Text task, the optimal performance also corresponds to a learning rate of 2e-5.
When the learning rate exceeds 2e-5, metrics decline slightly.

These results indicate that UniCA is most stable and effective at a learning rate of 2e-5.
A smaller learning rate leads to insufficient parameter updates, while a larger learning rate may cause overshooting in gradient descent, weakening the model’s ability to capture cross-modal correlations.
Thus, this paper selects 2e-5 as the optimal learning rate for subsequent experiments.

\section{Conclusion}
This study addresses the limitations of implicit cross-modal fusion in existing multi-modal retrieval frameworks by proposing UniCA, a Unified Cross-Attention Encoder model.
Through three core innovations—bi-directional cross-attention (Bi-CA) blocks, a Positive Similarity Loss, and the UMR-S10 lightweight dataset—UniCA achieves significant improvements in multi-modal semantic alignment and retrieval performance.

Experimental results on the WebQA benchmark validate the efficacy of our design.
UniCA consistently outperforms the VISTA-based baseline across both Hybrid and Image-Text retrieval tasks.
Ablation analyses further clarify our core contributions:
(1) The Bi-CA module enables explicit visual-textual token interaction, which is the primary driver of performance gains.
(2) The Positive Similarity Loss enhances embedding alignment by optimizing absolute semantic proximity, compensating for the relative-ranking bias of traditional cross-entropy loss.
(3) The UMR-S10 dataset reduces training costs while retaining most of the full-dataset performance, lowering barriers for resource-constrained scenarios.

Future work will focus on two targeted directions addressing existing limitations: expanding the UMR-S10 dataset by incorporating more diverse modalities (e.g., audio, video clips) alongside existing visual-textual data to support universal multi-modal retrieval scenarios covering heterogeneous information types, and investigating the impact of Bi-CA module layer count on performance—adjusting the number of cross-attention layers balances model complexity and cross-modal interaction quality, optimizing efficiency for lightweight deployment.
These refinements will further enhance UniCA’s adaptability and push the boundaries of efficient multi-modal retrieval systems.

\bibliographystyle{IEEEtran}
\nocite{*}
\bibliography{references}

\end{document}